\documentclass[runningheads]{llncs}

\usepackage{amssymb}
\usepackage{amsfonts}
\usepackage{amsmath}
\usepackage{cite}
\usepackage{color}
\usepackage{graphicx}
\usepackage[colorlinks=true,citecolor=blue,linkcolor=blue,urlcolor=blue]{hyperref}
\usepackage{cleveref}
\usepackage{pifont}
\usepackage{todonotes}
\usepackage{url}
\usepackage{xspace}

\graphicspath{ {./images/} }

\newcommand\para[1]{\subsubsection*{#1.}}
\newcommand\GG{\mathbb{G}}

\newcommand\HH{\mathsf{H}}
\newcommand\sk{\mathsf{sk}}
\newcommand\pk{\mathsf{pk}}

\newcommand\one{\ding{202}\xspace}
\newcommand\two{\ding{203}\xspace}
\newcommand\three{\ding{204}\xspace}

\ifdefined\publish
    \newcommand{\philipp}[1]{}
\else
    \newcommand{\philipp}[1]{\todo[inline, color=teal!20]{\textbf{Philipp:} #1}}
\fi

\begin{document}

\title{Performance of EdDSA and BLS Signatures in Committee-Based Consensus}

\author{
  Zhuolun Li\inst{1} \and
  Alberto Sonnino\inst{2,3} \and
  Philipp Jovanovic\inst{3}
}
\authorrunning{Z. Li et al.}

\institute{
  University of Leeds \and
  Mysten Labs \and
  University College London (UCL)
}

\maketitle

\begin{abstract}
  We present the first performance comparison of EdDSA and BLS signatures in committee-based consensus protocols through large-scale geo-distributed benchmarks.
  Contrary to popular beliefs, we find that \emph{small} deployments (less than 40 validators) can benefit from the small storage footprint of BLS multi-signatures while larger deployments should favor EdDSA to improve performance.
  As an independent contribution, we present a novel way for committee-based consensus protocols to verify BLS multi-signed certificates by manipulating the aggregated public key using pre-computed values.

\end{abstract}

\section{Introduction}

Consensus protocols run at the core of blockchains to order clients' transactions into a sequence agreed by all honest validators. The popularity of blockchains raised the interest in developing high-performance consensus systems, with early studies proposing committee-based protocols to improve over Bitcoin's~\cite{bitcoin} low throughput of 7 transactions per second.
These protocols have since been shown to increase blockchain throughput and reduce latency~\cite{DBLP:journals/corr/Kokoris-KogiasJ16, sok-consensus}, and they are rapidly becoming the standard in recent proof-of-stake architectures~\cite{diem, aptos, sui}. The blockchain literature provides a large variety of efficient committee-based consensus protocols. They improve on the state-of-the-art through various techniques, ranging from efficient data-sharing layers~\cite{narwhal,dispersedledger} and robust DAG-based protocols~\cite{bullshark,dumbo-ng} to multi-mode protocols adapting to faults and network conditions~\cite{gelashvili2021jolteon, dumbo-transformer, zyzzyva, 700bft}.

Unfortunately, these works pay little attention to their choice of signature scheme. Our inspection of their codebases indicates that most use either EdDSA~\cite{narwhal, bullshark, diem, aptos, sui} or BLS~\cite{dumbo-ng, dumbo-transformer, guo2020dumbo, guo2022speeding} but do not justify their choice. This is unfortunate as digital signatures require CPU-intensive operations and are extensively used in committee-based consensus: block proposers authenticate their block proposals by signing them, validators counter-sign block proposals to indicate their support, and certificates containing a quorum of signatures are used to commit and finalize transactions.
On the one hand, EdDSA signatures provide very fast signature generation and verification; on the other hand, BLS multi-signing enables small certificates and nearly constant-time certificate verification regardless of the committee size. It is a popular belief that BLS is preferable to EdDSA for large deployments where large EdDSA certificates are slow to propagate and verify. There is however no empirical evidence supporting this belief.

We address this gap by providing the first performance comparison of EdDSA and BLS signatures in committee-based consensus (to the best of our knowledge). We demonstrate through large-scale geo-distributed benchmarks that the choice of the signature scheme is a major factor determining the system's performance.
We find that contrary to popular beliefs, deployments with a relatively \emph{small} committee size (less than 40 validators) can benefit from the small storage footprint of BLS multi-signatures while larger deployments should favor EdDSA to improve performance. In a nutshell, the computational overhead of BLS verification becomes prohibitive when validators verify a large number of counter-signed block proposals; at the point where it offsets the benefits of small and efficient certificates. We select HotStuff~\cite{hotstuff} as an example of a committee-based consensus protocol for our experiments.
As an independent contribution, we present a novel way for committee-based consensus protocols to verify BLS multi-signed certificates by pre-computing a fixed number of group elements to manipulate the aggregated public key. This technique outperforms a traditional BLS verification process (even in the presence of Byzantine faults) requiring to re-compute the appropriate aggregated public key upon each certificate verification.

\para{Contributions}
In summary, we make the following contributions:
\begin{itemize}
  \item We perform the first performance comparison of EdDSA and BLS signatures in committee-based consensus (to the best of our knowledge) through large-scale geo-distributed benchmarks.
  \item We analyze the performance implications of each signature scheme and identify that small deployments are most suited to take advantage of BLS multi-signatures while larger deployments should favor EdDSA.
  \item We present a novel and more efficient way for committee-based consensus protocols to take advantage of BLS multi-signatures to verify certificates.
\end{itemize}

\section{Background} \label{sec:background}
We recall BLS multi-signatures and provide an overview of typical committee-based consensus protocols.

\para{BLS multi-signatures}
The Boneh-Lynn-Shacham (BLS) signature scheme~\cite{bls} is an efficient signature scheme using pairing-friendly elliptic curves. BLS supports multi-signing and public-key aggregation, making it very popular for various blockchain projects.
We start by recalling the standard BLS signature scheme. Let $\GG_1$, $\GG_2$, and $\GG_T$ be groups of prime order $q$ such that there exists an efficiently computable and non-degenerate bilinear map $e: \GG_1 \times \GG_2 \rightarrow \GG_T$. We denote by $g_1$, $g_2$, and $g_T$ the canonical generators of $\GG_1$, $\GG_2$, and $\GG_T$, respectively, and let $\HH_1 : \{0,1\}^* \to \GG_1$. We denote the security parameter by $\lambda$ and $\xleftarrow{\$}$ denotes sampling uniformly at random.
The BLS signature scheme consists of the following algorithms:
\begin{itemize}
  \item \textbf{\textsf{BLS.Setup}}$(1^{\lambda})$: Setup and output a bilinear group $par=(q, \GG_1, \GG_2, \GG_T, e, g_1, g_2)$.
  \item \textbf{\textsf{BLS.KeyGen}}$(par)$: Given the parameters $par$, output a pair of public/secret keys $(\pk, \sk)$ where $\sk \xleftarrow{\$} \mathbb{Z}_{q}^*$ and $\pk= g_2^{\sk} \in \GG_2$.
  \item \textbf{\textsf{BLS.Sign}}$(par, \sk,m)$: Given a message $m \in \{0,1\}^*$, output a signature $\sigma = \HH_1(m)^\sk \in \GG_1$.
  \item \textbf{\textsf{BLS.Verify}}$(par, \pk, \sigma, m)$: Given a public key $\pk \in \GG_2$, a signature $\sigma \in \GG_1$ and a message $m \in \{0,1\}^*$, output $1$ if $e(\sigma, g_2) = e(\HH_1(m), \pk)$ and $0$ otherwise.
\end{itemize}

\noindent BLS signatures can support multi-signing with public key aggregation. A multi-signature scheme (MSP) allows $n$ signers to generate a short signature $\sigma$, on the \emph{same} message $m$ (the size of the signature is independent of the number of signers). To verify the multi-signature one needs all the signer's public keys aggregated into a single key $apk$, $m$, and $\sigma$.

\begin{itemize}
  \item \textbf{\textsf{MSP.Setup}}$(1^{\lambda})$: Output \textbf{\textsf{BLS.Setup}}$(1^{\lambda})$.

  \item \textbf{\textsf{MSP.KeyGen}}$(par)$: Output \textbf{\textsf{BLS.KeyGen}}$(par)$.

  \item \textbf{\textsf{MSP.Sign}}$(par,\sk, m)$: Output \textbf{\textsf{BLS.Sign}}$(par, \sk, m)$.

  \item \textbf{\textsf{MSP.SigAggr}}$(\{\sigma_0,\dots,\sigma_n\})$: Output $\sigma=\prod_{i=1}^n \sigma_i$.

  \item \textbf{\textsf{MSP.KeyAggr}}$(par, \{\pk_1,\dots,\pk_n\})$: Output $apk=\prod_{i=1}^n \pk_i$.

  \item \textbf{\textsf{MSP.Verify}}$(par,apk,\sigma,m)$: Output \textbf{\textsf{BLS.Verify}}$(par, apk, \sigma, m)$.
\end{itemize}

\noindent An MSP scheme is usually vulnerable to rogue key attacks if the signers do not prove their ownership of the corresponding secret keys~\cite{boneh2018compact}. This is not a problem for committee-based blockchains because validator prove knowledge of their secret key $\sk$ by signing a special transaction before being eligible to join a future committee.

We note that threshold signatures are a generalization of multi-signatures requiring only a threshold $t \leq n$ of signatures (rather than $t=n$) to compute $\sigma$. Threshold signatures require complex distributed key generation mechanisms to generate the key pair of every signer when there is no natural trusted third party to run the setups protocol~\cite{gennaro1999dkg,kate2012distributed}. Furthermore, they do not extend naturally to settings where validators have different voting powers and are thus unsuitable for consensus protocols. As a result, we do not consider them in this paper as no committee-based blockchains use them within their consensus protocol (to the best of our knowledge).

\para{Committee-based consensus}
Committee-based blockchains typically divide time into a sequence of epochs (lasting roughly a day~\cite{sui, aptos, tendermint}). They elect a committee of $n=3f+1$ validators for each epoch (usually through proof-of-stake~\cite{sui, aptos}), where $f$ is the maximum number of faulty validators that the system can tolerate. The elected committee then `extends' the blockchain by sequencing clients' transactions using a Byzantine fault tolerant (BFT) protocol.

\begin{figure}[t]
  \centering
  \includegraphics[width=0.7\textwidth]{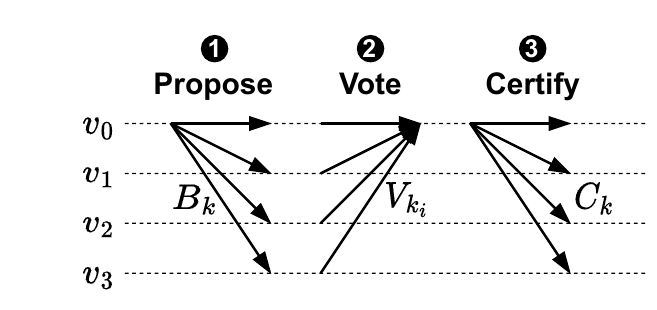}
  \caption{High-level overview of one round of a typical committee-based consensus protocol. The committee is formed of four validators $(v_0,v_1,v_2,v_3)$. The leader $v_0$ proposes $B_k$ for round $k$; validators reply with a vote $V_k$ over $B_k$; the leader collects a quorum of votes into a certificate $C_k$ and disseminates it, and validators verify $C_k$.}
  \label{fig:hotstuff}
\end{figure}

For the sake of this paper, we only present the aspects of committee-based consensus protocols where signatures intervene the most (and omit other aspects such as synchronizers~\cite{cohen2022proof}, leader-election modules~\cite{cohen2022aware}, and view-changes~\cite{castro1999practical,gelashvili2021jolteon}).
Typical committee-based consensus protocols operate in a round-by-round manner, electing a leader in each round among the validators to balance validator participation. \Cref{fig:hotstuff} provides a high-level overview of one round of a typical committee-based consensus protocol running with four validators, $v_0$, $v_1$, $v_2$, and $v_3$.
The leader $v_0$ of round $k$ disseminates a block $B_k$ extending the longest chain of blocks it knows\footnote{Usually leaders collect batches of transactions to propose, referred to as blocks, hence the protocol forms a chain of blocks (or a `blockchain').}~(\one). Validators then vote for at most one leader's proposal for each round by counter-signing it unless the proposal is malformed or conflicts with a longer chain that they know~(\two); validators send their votes $V_k$ back to the leader. The leader aggregates a quorum of $2f+1$ votes into a certificate $C_k$ and distributes it to the validators; validators accept $C_k$ if it is correctly signed by a quorum~(\three).

The protocol then repeats for several rounds (usually two or three) in order to commit $B_k$. For instance, the original HotStuff protocol~\cite{hotstuff} commits $B_k$ when there exist three consecutive certified blocks in the chain, $C_{k}$, $C_{k+1}$, $C_{k+2}$. More recent variants of HotStuff, such as Jolteon~\cite{gelashvili2021jolteon}, only require two consecutive rounds; and state-of-the-art DAG-based protocols~\cite{narwhal,bullshark,dumbo-ng} allow multiple validators to disseminate proposals in parallel. The general protocol flow, however, remains similar. If the leader fails or is unresponsive for a long periods of time, the validators run a \emph{view-change} sub-protocol to elect a new leader~\cite{castro1999practical}; changing leader is expensive and severely degrades performance.

The key motivation for BLS multi-signatures in committee-based consensus is to reduce the size of certificates (which grow linearly with the committee size) and allow their nearly constant-time verification (by verifying a single multi-signature rather than the $2f+1$ votes individually).

\section{Cached BLS Multi-Signature Verification} \label{sec:design}

We extend the BLS multi-signature scheme presented in \Cref{sec:background} with a new function \textbf{\textsf{MSP.KeyDisAggr}} that subtracts a set of public keys from the aggregate public key $apk$.
\begin{itemize}
    \item \textbf{\textsf{MSP.KeyDisAggr}}$(par, apk, \{-\pk_1,\dots,-\pk_n\})$: Given the aggregate public key $apk$, the opposite of all public keys $PK = \{-\pk_1,\dots,-\pk_n\}$, output $apk^* = apk \prod_{i=1}^j -\pk_i$.
\end{itemize}

This function requires a single elliptic curve addition per key to remove from the aggregate.
We now show how to incorporate it in the normal flow of committee-based consensus protocols depicted in \Cref{fig:hotstuff} of \Cref{sec:background}.

\para{Protocol description}
Every validator in the committee is initialized with the public parameters $par$ output by \textbf{\textsf{MSP.Setup}}$(1^{\lambda})$. Each validator locally runs \textbf{\textsf{MSP.KeyGen}}$(par)$ to generate their public/secret keypair $(\pk,\sk)$ and publishes $\pk$ (see \Cref{sec:background}).
Each validator stores the public key $\{\pk_1,\dots,\pk_n\}$ of all other valiators. They also compute and store the aggregated public key $apk = \textbf{\textsf{MSP.KeyAggr}}(par,\{\pk_1,\dots,\pk_n\})$ as well as the opposite of all validator's public keys $\{-\pk_1,\dots,-\pk_n\}$. 

\begin{itemize}
  \item \textbf{Step \one: Propose.} The leader of round $k$ collects a set of clients' transactions $l$ and creates a block proposal $m=(k,l,\cdot)$, where the dot `$\cdot$' denotes omitted protocol-specific fields. The leader then signs $m$ by calling $\sigma_B = \textbf{\textsf{MSP.Sign}}(par,\sk,m)$ using its secret key $\sk$ and disseminates $B_k(m, \sigma_B)$ to the other validators.

  \item \textbf{Step \two: Vote.} Validators first parse $B_k=(m, \sigma_B)$ and then verify it via $\textbf{\textsf{MSP.Verify}}(par,\pk,\sigma_B,m)$ where $\pk$ is the leader's public key. If the check passes and all other protocol-specific conditions are met, they counter-sign $B_k$ and send their vote $V_k = \textbf{\textsf{MSP.Sign}}(par,\sk, H(B_k))$ to the leader (where $H$ is a collision-resistant hash-function).

  \item \textbf{Step \three: Certify.} The leader calls $\textbf{\textsf{MSP.Verify}}(par,\pk,V_k,H(B_k))$, where $\pk$ is the public key of the voter, to verify each incoming vote. As soon as it receives $2f+1$ valid votes $\{V_{k_{1}},\dots,V_{k_{n}}\}$, it aggregates them calling $\sigma_C = \textbf{\textsf{MSP.SigAggr}}(\{V_{k_{1}},\dots,V_{k_{n}}\})$. It then computes a bitmap $b$ indicating which validators \emph{did not} contribute to $\sigma_C$. This is achieved by deterministically attributing an index to each validator that corresponds to its position in the bitmap. This bitmap allows to reduce the size of the certificate that would otherwise contain the public key of each signer. The leader then disseminates the certificate $C_k = (\sigma_C, b)$ to the validators.
        Upon receiving $C_k$, validators use the bitmap $b$ to identify the validators who did not contribute to $\sigma_C$, retrieve their previously cached set of $\{-\pk_1,\dots,-\pk_j\}$, and compute $apk^* = \textbf{\textsf{MSP.KeyDisAggr}}(par, apk, \{-\pk_1,\dots,-\pk_n\})$. They then verify the certificate calling $\textbf{\textsf{MSP.Verify}}(par,apk^*,\sigma_C,H(B_k))$.
\end{itemize}

This approach has two main advantages. (i) The bitmap allows to reduce the certificate size by 32B per signer compared to straightforward implementations including the public key of each signer in the certificate (as it is the case in many production systems~\cite{diem,aptos,sui}); certificates are part of the forever-stored blockchain so any message compression becomes substantial over time. (ii) Before verifying a certificate validators compute $apk^*$ with at most $f$ elliptic curve additions, while a straightforward BLS multi-signature verification would require $2f+1$ to recompute the aggregated verification key every time. Furthermore, practical leaders' implementations wait around 50-100ms after collecting the first $2f+1$ votes to give extra time to the remaining validators to vote. As a result, typical certificates contain close to $n=3f+1$ votes in the common case (happy path), and computing $apk^*$ requires only one or two elliptic curve additions.

\section{Performance Comparison} \label{sec:performance}
We select HotStuff~\cite{hotstuff} as an example of a committee-based consensus protocol for our experiments. We select this protocol because it is the quorum-based consensus protocol most used in production blockchains; Celo~\cite{celo}, Cypherium~\cite{cypherium}, Flow~\cite{flow}, Diem~\cite{diem}, and Aptos~\cite{aptos} all run a variant of HotStuff. Furthermore, it shares many design traits with Tendermint~\cite{tendermint} (its closest ancestor). We specifically run our benchmarks on a 2-chain HotStuff variant called Jolteon~\cite{gelashvili2021jolteon}; we chose this variant because Diem~\cite{diem}, Aptos~\cite{aptos}, and Flow~\cite{flow} run it in production.

\para{Implementation}
We implement BLS multi-signatures on top of the original open-source implementation of Jolteon\footnote{
  \url{https://github.com/asonnino/hotstuff}
}.
It is implemented in Rust, uses Tokio\footnote{\url{https://tokio.rs}} for asynchronous networking, ed25519-dalek\footnote{\url{https://github.com/dalek-cryptography/ed25519-dalek}} for signatures, and data-structures are persisted using RocksDB\footnote{\url{https://rocksdb.org}}. It uses TCP to achieve reliable point-to-point channels, necessary to correctly implement the distributed system abstractions.
Since this implementation uses EdDSA (over the curve Ed25519), we modify its \texttt{crypto} module to use BLS multi-signatures as described in \Cref{sec:design}. We use the BLS implementation of Filecoin (over the curve BLS12-381)\footnote{
  \url{https://github.com/filecoin-project/bls-signatures}
}.
We are open-sourcing our BLS-enabled implementation\footnote{
  \url{https://github.com/radiken/hotstuff-digital-signature-benchmarking}
}.

\para{Evaluation setup}
We evaluate the throughput and latency of HotStuff/Jolteon equipped with BLS multi-signatures through experiments on Amazon Web Services (AWS). We then compare its performance with the baseline implementation using EdDSA for various committee sizes.

We deploy a testbed on AWS, using \texttt{t3.medium} instances across 4 different AWS regions: N. Virginia (us-east-1), N. California (us-west-1), Sydney (ap-southeast-2), and Frankfurt (eu-central-1). Validators are distributed across those regions as equally as possible. Each machine provides up to 5 Gbps of bandwidth, 2 virtual CPUs (1 physical core) on a 2.5 GHz, Intel Xeon Platinum 8175, 4 GB memory, and runs Linux Ubuntu server 20.04.
In the following sections, each measurement in the graphs is the average of 5 independent runs, and the error bars represent one standard deviation.
Our baseline experiment parameters are a block size of 500KB, a transaction size of 512B, and one benchmark client per party submitting transactions at a fixed rate for 5 minutes. The leader timeout value is set to 5 seconds. When referring to \emph{latency}, we mean the time elapsed from when the client submits the transaction to when the transaction is committed by one validator. We measure it by tracking sample transactions throughout the system.

\para{Analysis}

\begin{figure}[t]
  \begin{minipage}{0.49\textwidth}
    \centering
    \includegraphics[width=\textwidth]{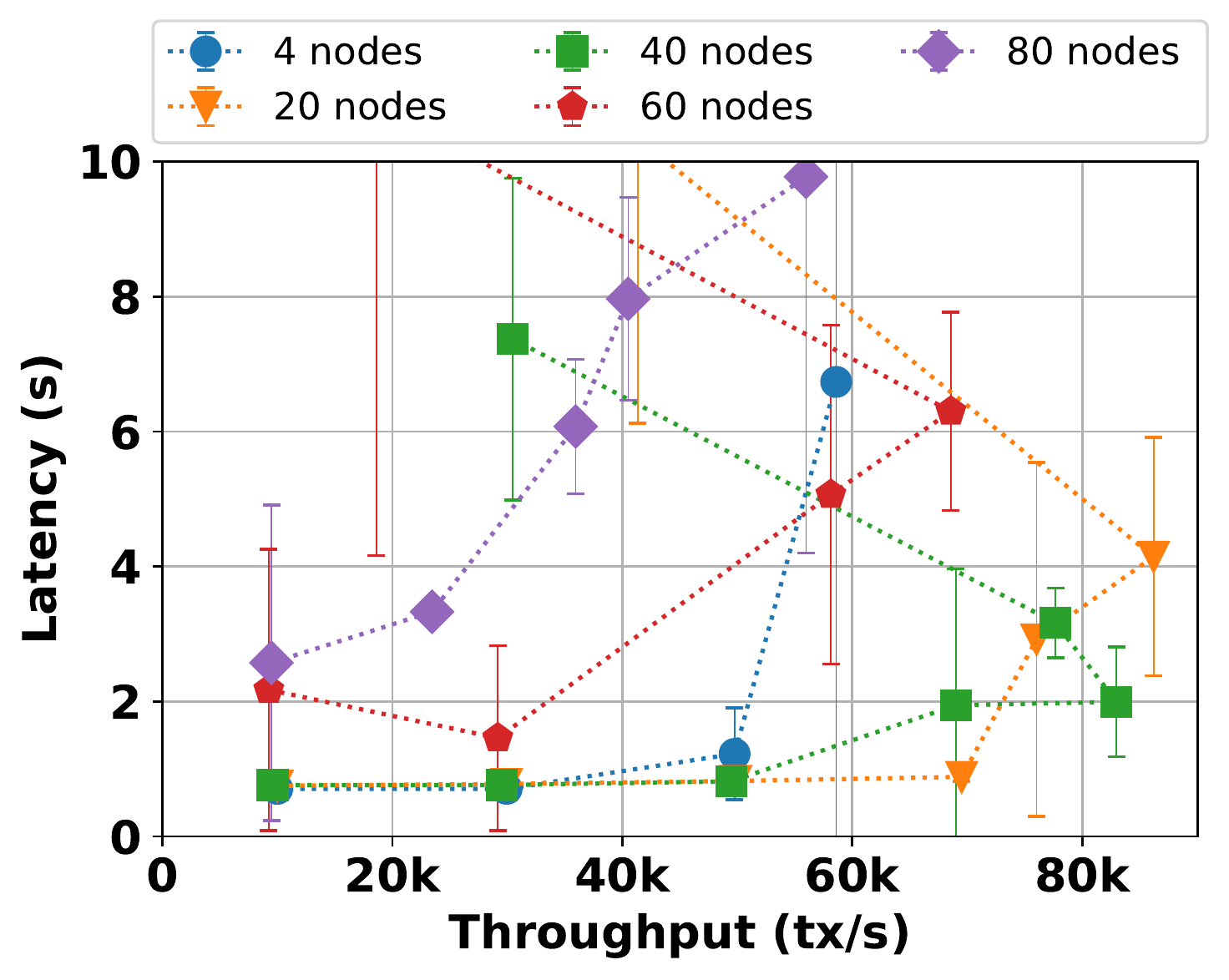}
    \caption{EdDSA-based implementation; 4, 20, 40, 60, and 80 validators (WAN).}
    \label{fig:eddsa}
  \end{minipage}
  \centering
  \begin{minipage}{0.49\textwidth}
    \centering
    \includegraphics[width=\textwidth]{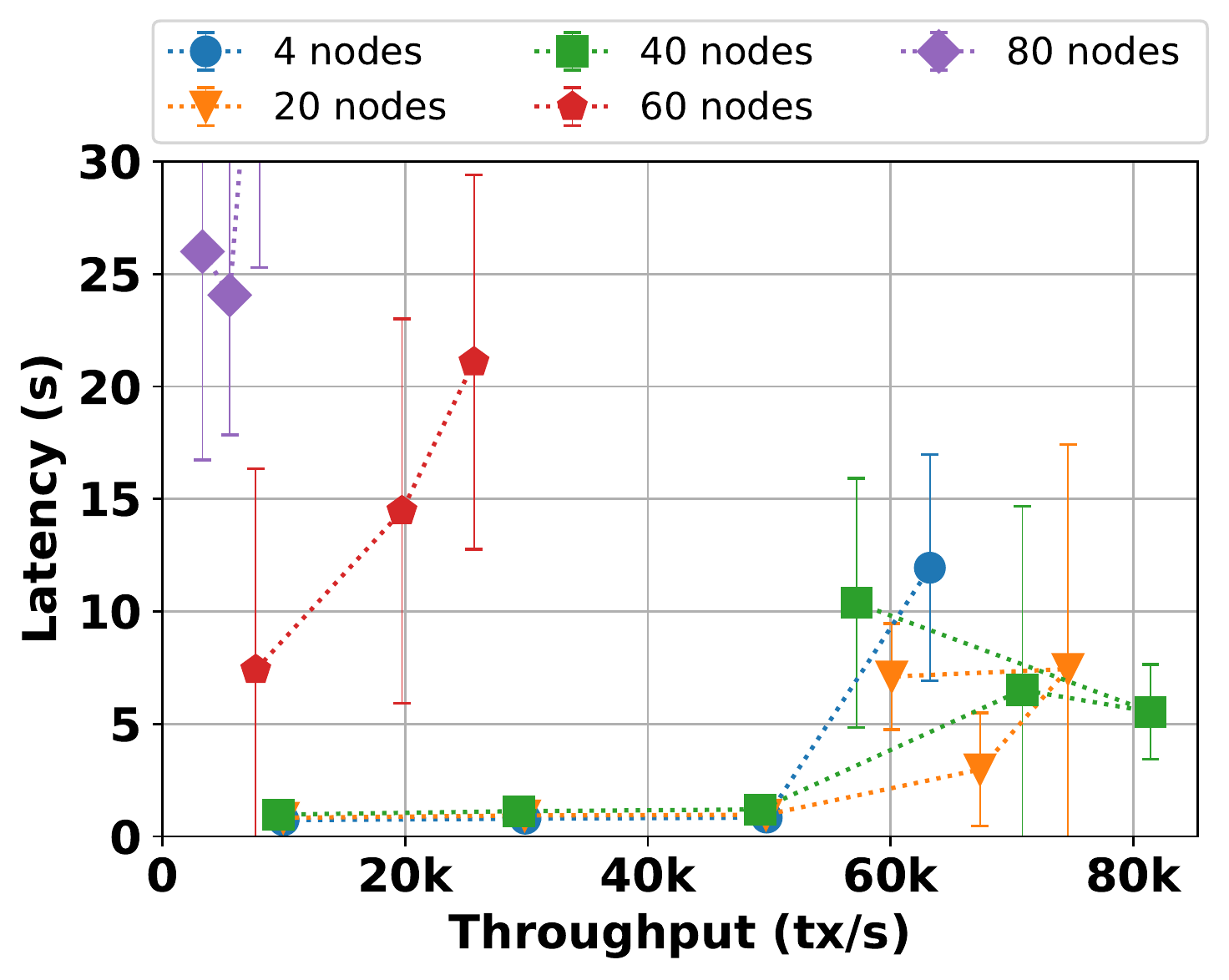}
    \caption{BLS-based implementation; 4, 20, 40, 60, and 80 validators (WAN).}
    \label{fig:bls}
  \end{minipage}
\end{figure}

\Cref{fig:eddsa} and \Cref{fig:bls} show that a 4-validators deployment can process 50,000 tx/s while keeping the latency below 1.5 seconds, regardless of the signature scheme. Similarly, 20- and 40-validator deployments can process around 80,000 tx/s while keeping the latency around 4 to 5 seconds, regardless of the signature scheme\footnote{It may seem surprising that the system achieves a higher throughput with 20 and 40 validators than with 4. This is however a known result~\cite{narwhal,gelashvili2021jolteon}, the extra capacity provided by the additional validators allows for better resource utilization.}.

Increasing the committee size to 60 validators drops the performance of our EdDSA-based implementation to around 60,000 tx/s and increases the latency to 5 seconds (\Cref{fig:eddsa}). This performance drop is explained by both the additional bandwidth required to broadcast messages to many validators and the CPU overhead required to verify a large number of signatures. Indeed, the leader needs to verify at least $2f+1=41$ votes every round (step~\two of \Cref{fig:hotstuff}) and validators need to verify 41 signatures to validate each certificate\footnote{The `batch-verify' feature of EdDSA greatly speeds up certificate verification.} (step~\three of \Cref{fig:hotstuff}).
Our BLS-based implementation suffers a more significant performance drop: \Cref{fig:bls} indicates the system can only process up to 20,000 tx with a latency of about 15 seconds. It appears that this performance difference is due to the time required by the leader to verify individual votes (step~\two of \Cref{fig:hotstuff}). EdDSA allows the leader to efficiently verify votes while BLS verification is about 100x slower and monopolizes the leader's CPU. This causes the leader's slow down, which affects both throughput and latency.

\Cref{fig:eddsa} indicates that even larger deployments of 80 validators further drop the performance of the EdDSA-based implementation to about 40,000 tx/s (with a latency of 6 seconds). \Cref{fig:bls} shows that our BLS-based implementation barely manages to process transactions; it can only process a few thousands tx/s with a latency of over 25 seconds. The time required by the leader to verify votes exceeds 5 seconds most of the time (the leader-timeout value), at which point validators believe the leader crashed and initiate a view-change sub-protocol to elect a new leader; this scenario repeats often andgreatly degrades performance.

\para{Key takeaways}
Our experiments demonstrate no apparent performance benefit of BLS signatures. Contrary to popular belief, large deployments do not benefit from the aggregation properties of BLS. Despite BLS multi-signatures enabling small certificates and nearly constant-time certificate verification (regardless of the committee size), the CPU overhead of individual votes verification greatly offsets this benefit. As a result, large deployments should favor EdDSA.
Small deployments (up to 40 validators) do not place an excessive CPU burden on the leader, EdDSA and BLS-based deployments perform similarly and they may thus take advantage of the aggregation properties of BLS. Small BLS multi-signed certificates can provide a significant storage benefit: storing an EdDSA certificate requires 2.5 KB (for a 40 nodes deployment) while a BLS multi-signed certificate only requires about 100 B. This difference may become significant over time since certificates are part of the forever-persisted blockchain.

\section*{Acknowledgements}
This work is partially funded by Mysten Labs.

\bibliographystyle{splncs04}
\bibliography{references}
\end{document}